\documentclass[11pt,a4paper]{article}
\usepackage[utf8]{inputenc}

\usepackage{jheppub}
\usepackage{amsmath}
\usepackage{breqn}
\usepackage{comment}
\usepackage[colorinlistoftodos]{todonotes}
\usepackage{float}
\usepackage{graphics}

\graphicspath{{Figure/}}
\usepackage{caption} 
\usepackage{subcaption} 
\usepackage{amsfonts}
\usepackage{todonotes} 

\title{Angular phase-space integrals with four denominators through Mellin--Barnes} 

\author[a]{Taushif Ahmed,}
\author[a]{Syed Mehedi Hasan,}
\author[a]{Andreas Rapakoulias}
\emailAdd{taushif.ahmed@ur.de}
\emailAdd{Syed-Mehedi.Hasan@ur.de}
\emailAdd{andreas.rapakoulias@ur.de}

\affiliation[a]{Institute for Theoretical Physics, University of Regensburg, 93040 Regensburg, Germany}
\preprint

\abstract{
We compute four-denominator angular phase-space integrals using the Mellin--Barnes (MB) technique in dimensional regularisation. Independent of the scattering process, an angular integral can be categorised based on the nature of the momenta appearing in the denominators. We address all scenarios involving fully massless and massive momenta. We present a partial fraction decomposition that relates angular integrals with multiple massive momenta to those with a single massive momentum. By solving six- and seven-fold MB integrals, we express the final results up to the finite order in the dimensional 
regulator in terms of Goncharov polylogarithms. 
}

\newcommand{\ep}{\epsilon}

\newcommand{\be}{\begin{equation}}
\newcommand{\ee}{\end{equation}}
\newcommand{\bea}{\begin{eqnarray}}
\newcommand{\eea}{\end{eqnarray}}

\definecolor{Red}{rgb}{1.,0.,0.}
\definecolor{randomcolour}{rgb}{0.2,0.5,0.7}

\DeclareMathAlphabet\mathbfcal{OMS}{cmsy}{b}{n}

\def\OMIT#1{}

\definecolor{darkred}{rgb}{0.9,0,0}

\definecolor{darkgreen}{rgb}{0,0,0.9}

\definecolor{darkblue}{rgb}{0,0,0.9}

\usepackage{cancel}
\allowdisplaybreaks[1]

\begin{document}
\allowdisplaybreaks[4]
\unitlength1cm
\maketitle
\flushbottom


\def\D{{\cal D}}
\def\DD{\overline{\cal D}}
\def\g{\overline{\cal C}}
\def\gm{\gamma}
\def\M{{\cal M}}
\def\ep{\epsilon}
\def\epm1{\frac{1}{\epsilon}}
\def\epm2{\frac{1}{\epsilon^{2}}}
\def\epm3{\frac{1}{\epsilon^{3}}}
\def\epm4{\frac{1}{\epsilon^{4}}}
\def\unM{\hat{\cal M}}
\def\ashat{\hat{a}_{s}}
\def\asmur{a_{s}^{2}(\mu_{R}^{2})}
\def\sigbar{{{\overline {\sigma}}}\left(a_{s}(\mu_{R}^{2}), L\left(\mu_{R}^{2}, m_{H}^{2}\right)\right)}
\def\sigbarn{{{{\overline \sigma}}_{n}\left(a_{s}(\mu_{R}^{2}) L\left(\mu_{R}^{2}, m_{H}^{2}\right)\right)}}
\def\unas{ \left( \frac{\hat{a}_s}{\mu_0^{\epsilon}} S_{\epsilon} \right) }
\def\rnM{{\cal M}} 
\def\bt{\beta}
\def\cD{{\cal D}}
\def\cC{{\cal C}}
\def\ca{\text{\tiny C}_\text{\tiny A}}
\def\cf{\text{\tiny C}_\text{\tiny F}}
\def\ct{{\red []}}
\def\sv{\text{SV}}
\def\murOmu{\left( \frac{\mu_{R}^{2}}{\mu^{2}} \right)}
\def\bb{b{\bar{b}}}
\def\bt0{\beta_{0}}
\def\bt1{\beta_{1}}
\def\bt2{\beta_{2}}
\def\bt3{\beta_{3}}
\def\gm0{\gamma_{0}}
\def\gm1{\gamma_{1}}
\def\gm2{\gamma_{2}}
\def\gm3{\gamma_{3}}
\def\nn{\nonumber}
\def\l{\left}
\def\r{\right}
\def\T{{\cal Z}}    
\def\U{{\cal Y}}

\def\nn{\nonumber\\}
\def\ep{\epsilon}
\def\T{\mathcal{T}}
\def\V{\mathcal{V}}

\def\qgraf{{\fontfamily{qcr}\selectfont
QGRAF}}
\def\python{{\fontfamily{qcr}\selectfont
PYTHON}}
\def\form{{\fontfamily{qcr}\selectfont
FORM}}
\def\reduze{{\fontfamily{qcr}\selectfont
REDUZE2}}
\def\kira{{\fontfamily{qcr}\selectfont
Kira}}
\def\litered{{\fontfamily{qcr}\selectfont
LiteRed}}
\def\fire{{\fontfamily{qcr}\selectfont
FIRE5}}
\def\air{{\fontfamily{qcr}\selectfont
AIR}}
\def\mint{{\fontfamily{qcr}\selectfont
Mint}}
\def\hepforge{{\fontfamily{qcr}\selectfont
HepForge}}
\def\arXiv{{\fontfamily{qcr}\selectfont
arXiv}}
\def\Python{{\fontfamily{qcr}\selectfont
Python}}
\def\ginac{{\fontfamily{qcr}\selectfont
GiNaC}}
\def\polylogtools{{\fontfamily{qcr}\selectfont
PolyLogTools}}
\def\anci{{\fontfamily{qcr}\selectfont
Finite\_ppbk.m}}

\newcommand{\dis}{}
\newcommand{\overbar}[1]{mkern-1.5mu\overline{\mkern-1.5mu#1\mkern-1.5mu}\mkern
1.5mu}
\section{Introduction}
The analytic computation of scattering cross sections at high perturbative orders in quantum field theory is a cornerstone of modern collider phenomenology. With the increasing precision of experimental data from facilities such as the Large Hadron Collider (LHC) and future colliders, such as the Electron-Ion Collider (EIC), there is a strong demand for equally precise theoretical predictions. Such predictions require the evaluation of both virtual and real-emission contributions at next-to-leading order (NLO), next-to-next-to-leading order (NNLO), and beyond. While the last two decades have witnessed tremendous advances in the analytic and numerical evaluation of multi-loop Feynman integrals for virtual amplitudes, the treatment of real-emission phase-space (PS) integrals—particularly their angular components—has progressed more slowly despite their equal physical importance.

Phase-space integrals in dimensional regularization factorize into radial and angular parts in a suitably chosen reference frame~\cite{Somogyi:2011ir,Anastasiou:2013srw}. While the radial part depends on a specific scattering process, the angular part is universal. This indicates that one can express the latter in a generic form. As it turns out, this can be recast into a form having an $n$ propagator-like denominators~\cite{Somogyi:2011ir}:  
\begin{align}
\label{eq:ang_int}
\Omega_{j_{1}, \ldots, j_{n}}(\{p_i^\mu\},d) \equiv \int \mathrm{d} \Omega_{d-1}(q) \,
\frac{1}{\left(p_{1} \cdot q\right)^{j_{1}} \ldots\left(p_{n} \cdot q\right)^{j_{n}}} \,,
\end{align}
where $\mathrm{d} \Omega_{d-1}(q)$ is the rotationally invariant measure for a massless vector $q^\mu$ in $d = 4 - 2\epsilon$ dimensions, $\{p_i^\mu\}$ are fixed reference momenta, and the integers $\{j_i\}$ depend on the process and perturbative order. The vectors $\{p_i^\mu\}$ are normalised to dimensionless. Given a scattering process, one has to identify or define $\{p_i^\mu\}$ in terms of the momenta and other parameters present in the system. 
Angular integrals are categorised according to the number of propagator denominators and the nature of the reference momenta $p_i^\mu$, whether massless or massive.  
An increase in either of these features quickly increases the complexity, much like in loop Feynman integrals.
Despite being a universal part, when combining the angular integral with the radial component, extra care must be taken due to the possible presence of soft divergences~\cite{Ahmed:2024pxr,Ahmed:2024owh}.

Such integrals~\cite{Schellekens:1981kq, vanNeerven:1985xr, Beenakker:1988bq, Somogyi:2011ir, Lyubovitskij:2021ges, Wunder:2024btq,Smirnov:2024pbj,Ahmed:2024pxr,Haug:2024yfi,Haug:2025sre} arise ubiquitously in higher-order QCD corrections. Depending on the number of real emission particles, which is governed by the perturbative order under consideration, the number of required denominators changes in \eqref{eq:ang_int}. To begin with, although there might be a larger number of propagators, utilising partial fractions, we can often reduce it. For example, in NNLO double-real emission in semi-inclusive deep inelastic scattering (SIDIS), the most complicated case needed involves two denominators~\cite{Anderle:2016kwa,Wang:2019bvb,Ahmed:2024owh}. 
The analytic structure of~\eqref{eq:ang_int} is well understood for a few simple cases:
\begin{itemize}
\item \textbf{$n=1$}: For a massless $p_1$, the integral reduces to the total angular volume. For massive $p_1$, the result is given by a Gauss hypergeometric function ${}_2F_1$~\cite{vanNeerven:1985xr,Somogyi:2011ir}.
\item \textbf{$n=2$}: With both $p_1$ and $p_2$ massless, the result again involves ${}_2F_1$~\cite{vanNeerven:1985xr,Somogyi:2011ir}. When one or both are massive, Appell $F_1$ or Lauricella functions appear~\cite{vanNeerven:1985xr,Beenakker:1988bq,Somogyi:2011ir,Lyubovitskij:2021ges,Wunder:2024btq}. In ref.~\cite{Lyubovitskij:2021ges,Ahmed:2024owh}, their solutions in powers of $\epsilon$ is discussed. In the latter, it is discussed in the context of NNLO SIDIS QCD computation.
\item \textbf{$n = 3$}: The angular integral for $n \ge 3$ has an exact representation to all orders in $\epsilon$ in terms of the multivariable $H$-function. While formally compact, this form is notoriously difficult to expand in $\epsilon$ for general kinematics. For the specific case $n=3$, refs.~\cite{Ahmed:2024pxr} and~\cite{Haug:2024yfi} compute the expansion up to ${\cal O}(\epsilon^2)$ and ${\cal O}(\epsilon)$, respectively. In the former, we presented the result in terms of Goncharov polylogarithms (GPLs), unlike the latter, where the results are expressed in terms of Clausen functions. More recently, ref.~\cite{Haug:2025sre} discusses the structure of the all-orders $\epsilon$ dependence.  In ref.~\cite{Smirnov:2024pbj}, the authors investigate multi-denominator integrals in the small-mass limit. 
\end{itemize}
\textit{In this work}, we employ a Mellin--Barnes (MB) representation to tackle $n=4$ integrals, enabling their expansion as a Laurent series in the dimensional regulator~$\epsilon$ for both massless and massive cases. The massless and single-massive 
cases are computed explicitly by evaluating the multi-fold MB integrals, whereas the other massive configurations are solved using a partial fraction decomposition. The latter was introduced in ref.~\cite{Lyubovitskij:2021ges} for the two-propagator case and extended to the three-propagator case in refs.~\cite{Ahmed:2024pxr,Haug:2024yfi}. In this article, we extend this method to the four-denominator case and present its explicit forms. In particular, we provide the result for $n=4$ using an MB-based method recently demonstrated in refs.~\cite{Ahmed:2024owh,Ahmed:2024pxr}, with the expression given 
in terms of GPLs up to $\mathcal{O}(\epsilon^{0})$. 
While this manuscript was in preparation, the same case was addressed in 
ref.~\cite{Haug:2025sre}\footnote{Ref.~\cite{Haug:2025sre} appeared online during the preparation of the present draft.} using the method of differential equations combined with dimensional shift relations. Unlike ours, the results are not given in terms of GPLs, but rather in terms of Clausen functions.\footnote{While refs.~\cite{Haug:2024yfi,Haug:2025sre} offer a compact representation of the three and four-denominator angular integrals in terms of the Clausen function; such a form is far less practical for combining with the parametric part to obtain the full phase-space integrals as fully analytic functions or as iterated integrals, where such a representation is attainable. By contrast, expressing the result in terms of GPLs is both natural and advantageous: it exploits their iterative structure, enabling the final result to be written in a fully analytic form, and it benefits from the availability of robust, well-tested numerical libraries~\cite{Vollinga:2004sn,Naterop:2019xaf}.  
The inevitable trade-off is that a GPL representation is unavoidably more extended in length--a consequence of analytic completeness rather than inefficiency. The numerical evaluation of the four-denominator case, including all massive denominators, completes almost instantaneously.} The single-massive four-denominator angular integral is computed up to 
$\mathcal{O}(\epsilon^{0})$ in ref.~\cite{Salvatori:2024nva} as a demonstration of a novel approach based on tropical geometry. 

This article is also significant in another regard: it presents the solution of six- and seven-fold MB integrals in terms of GPLs. To the best of our knowledge, this is the first time such high-fold MB integrals have been expressed in terms of analytic functions, such as GPLs, in the literature. This method is particularly powerful because it is both algorithmic and scalable. This achievement has a broad range of applications, extending well beyond the context of angular integrals.

Although the method naturally yields results in GPL form, GPLs do not 
necessarily span the complete functional space for every problem; the 
relevant space depends on the perturbative order and mass scales involved. When additional structures arise, the results can instead be cast in terms of iterative integrals, which exhibit many of the desirable properties and remain an efficient and versatile representation -- a situation familiar from the study of loop Feynman integrals.

The article is organised as follows: in section~\ref{sec:MB}, the MB representation of angular integrals and its expansion in Laurent series in $\epsilon$ is discussed. The resulting integrals can be converted into a set of multi-dimensional integrals over real parameters. This is discussed in section~\ref{sec:ana-int}. We solve these integrals in terms of GPLs. We schematically present the prescription in section~\ref{sec:toGPL}. In section~\ref{sec:4den-massless} and \ref{sec:masive}, we present the massless and massive cases. The partial fraction decomposition to express the double and higher mass angular integrals in terms of single massive ones is discussed in subsection~\ref{sec:double-massive}. The recursion relations that relate integrals with a higher number of propagators to the lower ones are discussed in section~\ref{sec:recursion}. We conclude in section~\ref{sec:concl} and present our findings in {\tt Mathematica} readable ancillary files.

\section{Mellin-Barnes representation of angular integrals}
\label{sec:MB}

The angular integral in Eq.~\eqref{eq:ang_int} can be cast into the Mellin–Barnes (MB) form
\begin{align}
\label{eq:MB}
\Omega_{j_{1}, \ldots, j_{n}}\left(\left\{v_{k l}\right\}, \epsilon\right)=&  \frac{2^{2-j-2 \epsilon} \pi^{1-\epsilon}}{\prod_{k=1}^{n} \Gamma\left(j_{k}\right) \Gamma(2-j-2 \epsilon)} 
  \int_{-\mathrm{i} \infty}^{+\mathrm{i} \infty}\bigg[\prod_{k=1}^{n} \prod_{l=k}^{n} \frac{\mathrm{d} z_{k l}}{2 \pi \mathrm{i}} \Gamma\left(-z_{k l}\right)\left(v_{k l}\right)^{z_{k l}}\bigg]\nonumber\\
\times &\bigg[\prod_{k=1}^{n} \Gamma\left(j_{k}+z_{k}\right)\bigg] \Gamma(1-j-\epsilon-z)\,,
\end{align}
where the scalar products among the vectors $p_i^\mu$ are encoded in
\begin{align}
\label{eq:sp}
v_{k l} \equiv \begin{cases}\frac{p_{k} \cdot p_{l}}{2}& ~{\rm for}~ k \neq l \\ \frac{p_{k}^{2}}{4} & ~{\rm for}~ k=l\end{cases}\,,
\end{align}
and the MB variables satisfy
\begin{align}
z &= \sum_{k=1}^{n} \sum_{l=k}^{n} z_{kl},
&
z_{k} &= \sum_{l=1}^{k} z_{lk} + \sum_{l=k}^{n} z_{kl}.
\end{align}
Thus, $z$ denotes the total sum over all MB variables, while $z_k$ collects all $z_{kl}$ involving index $k$, with $z_{kk}$ counted twice:
\[
z_k = z_{1k} + \dots + z_{k-1,k} + 2z_{kk} + z_{k,k+1} + \dots + z_{kn}.
\]
We also define $j = \sum_{k=1}^n j_k$.

Eq.~\eqref{eq:MB} holds for $\Re(j_k) > 0$; other values can be reached via analytic continuation. We further assume $v_{kl} \neq 0$. If $v_{kl} = 0$ for some $(k,l)$—as in the case of a massless leg with $v_{ii} = 0$—the corresponding $z_{kl}$ integration is omitted and $z_{kl}$ is set to zero in the remainder of the expression. The parameters lie in the range $v_{kl} \in (0,1)$ and they are also defined as dimensionless quantities as the vectors $p_i^\mu$ are.

In general, multifold MB integrals of the type \eqref{eq:MB} are not solvable in closed analytic form. Our interest is in expanding them in their Laurent expansion around $\epsilon = 0$, rather than evaluating them exactly. We proceed as follows:
\begin{enumerate}
    \item \textbf{Expansion around $\epsilon = 0$} \\
    The integrand of Eq.~\eqref{eq:MB} is formally expanded in a series in $\epsilon$. Each coefficient of the expansion still contains MB integrals, but typically with reduced complexity.
    
    \item \textbf{Analytic continuation} \\
    As $\epsilon \to 0$, poles from different Gamma functions may move and cross the original contours. In such cases, a direct expansion is ill-defined. Rather, an analytic continuation becomes necessary. This occurs in cases where the real parts of some gamma function arguments in the integrand take non-positive values.
    The integration contours in Eq.~\eqref{eq:MB} are taken to be parallel to the imaginary axis, chosen so that all poles of Gamma functions of the type $\Gamma(a+z)$ lie to the left, and all poles of the type $\Gamma(a-z)$ lie to the right.
    To handle contour crossings, we employ either:
    \begin{itemize}
        \item \emph{Contour deformation}~\cite{Smirnov:1999gc}: the contour is shifted in the real direction to restore pole separation; residues of crossed poles are added to the integral.
        \item \emph{Fixed-contour method}~\cite{Tausk:1999vh,Czakon:2005rk}: the contour is kept fixed while tracking pole motion with $\epsilon$; each time a pole crosses a contour, the corresponding residue is added.
    \end{itemize}
    
    \item \textbf{Iterative residue extraction} \\
    Residue terms may themselves involve MB integrals over the remaining variables. If additional contour crossings occur in these integrals as $\epsilon$ varies, the procedure is iterated until all resulting MB integrals are free of such singular configurations, at which point the integrand can be safely expanded. Thus, the original MB integral is effectively decomposed into several distinct MB integrals.

    \item \textbf{Final recombination} \\
    After all expansions are performed, contributions of the same order in $\epsilon$ from all MB pieces (original and residue terms) are combined to obtain the Laurent coefficients of the original integral.
\end{enumerate}

The primary objective is then to evaluate the MB integrals that appear as coefficients in the expansion.
Section~\ref{sec:ana-int} presents our strategy for evaluating the MB integrals and expressing the results in terms of multiple polylogarithms or, more generally, iterated integrals. For multifold MB integrals, however, a complete solution via this approach becomes considerably more involved. In section~\ref{sec:4den-massless}, we demonstrate these difficulties through the example of the four-denominator case.

\section{Conversion of MB integrals to real integrals}
\label{sec:ana-int}

In this section, we describe the procedure for analytically evaluating the MB integrals that arise as coefficients in the Laurent expansion around $\epsilon = 0$. Our approach leverages the concept of \emph{balanced} MB integrals, which allows a systematic reduction of the MB representation to integrals over real variables, facilitating their explicit evaluation.

Consider a one-dimensional MB integral of the form
\begin{equation}
\int_{-i\infty}^{+i\infty} \frac{dz_j}{2\pi i} \prod_{k=1}^{n_+} \Gamma(a_k + z_j)^{\alpha_k} \prod_{l=1}^{n_-} \Gamma(a_l - z_j)^{\beta_l},
\quad \alpha_k, \beta_l \in \mathbb{Z},
\end{equation}
where the integration contour is chosen such that the poles of $\Gamma(a_k + z_j)$ lie to the left and those of $\Gamma(a_l - z_j)$ lie to the right of the contour. We say this integral is \emph{balanced} in the variable $z_j$ if the integer exponents satisfy the condition
\begin{equation}
\sum_{k=1}^{n_+} \alpha_k = \sum_{l=1}^{n_-} \beta_l.
\end{equation}
For a multidimensional MB integral, the integrand depends on multiple variables $z_1$, $z_2$, \ldots, and the integral is balanced if this condition holds for each integration variable individually.

The significance of this balancing condition lies in the fact that the product of gamma functions in the integrand can then be rearranged into a product of Euler beta functions. This rearrangement is possible after performing analytic continuation and expanding around $\epsilon=0$ to ensure all gamma functions have positive real parts. Under these conditions, none of the gamma functions depend on the integration variables anymore, allowing their combination into beta functions:
\begin{equation}
B(a,b) = \frac{\Gamma(a)\Gamma(b)}{\Gamma(a+b)},
\end{equation}
where the parameters satisfy $\Re(a) > 0$, $\Re(b) > 0$.
The beta function admits integral representations that can be chosen depending on the problem context:
\begin{equation}
B(a,b) =
\begin{cases}
\displaystyle \int_0^1 dx x^{a-1} (1 - x)^{b-1}, \\\\
\displaystyle \int_0^\infty dx x^{a-1} (1 + x)^{-a - b}.
\end{cases}
\end{equation}
Because the integrals involved are convergent in the relevant domain, the order of integration can be exchanged, allowing us to bring the MB integrals inside the real integrals derived from the beta functions. 

Following this procedure, the original multidimensional MB integral can be recast into the form
\begin{equation}
\label{eq:real-mb}
\int_0^\kappa \left( \prod_{i=1}^K dx_i \right) R_0(\mathbf{x}, \mathbf{u}) \int_{-i\infty}^{+i\infty} \left( \prod_{l=1}^L \frac{d\tilde{z}_l}{2\pi i} \right) \left[ R_l(\mathbf{x}, \mathbf{u}) \right]^{\tilde{z}_l},
\end{equation}
where 
$\mathbf{x} = (x_1, \ldots, x_K)$ 
are real variables introduced by the beta function representations, and 
$\mathbf{u} = (u_1, \ldots, u_N)$ 
denote the external parameters and kinematic invariants present in the original MB integral but independent of the integration variables. The functions 
$R_0, R_l$ 
are rational expressions built from products and ratios of the variables 
$x_i$, 
terms of the form 
$(1 \pm x_i)$, 
and parameters 
$u_j$. 
The upper limit 
$\kappa$ 
is either 1 or 
$\infty$, 
depending on the choice of beta function representation.

Employing the identity
\begin{equation}
    \int^{+i\infty+z_{0}}_{-i\infty+z_{0}} \frac{dz}{2\pi i} A^{z} = \delta(1-A), \hspace{10pt} A>0,
\end{equation} 
the remaining MB integrals over $\tilde{z}_l$ 
can often be evaluated explicitly, transforming the problem into evaluating a finite set of integrals over real variables 
$\mathbf{x}$:
\begin{equation}
\int^{\kappa}_{0} \Bigg( \prod^{K}_{i=1} dx_{i} \Bigg) R_{0}(\mathbf{x},\mathbf{u}) \prod^{L}_{l=1} \delta \Bigg[1 - R_{l}(\mathbf{x},\mathbf{u})\Bigg].
\end{equation}
The resulting integrands are typically composed of rational functions and logarithms of polynomials in the 
$x_i$. 
When these integrals can be performed iteratively, the final results are expressible in terms of multiple polylogarithms (MPLs) or more general iterated integrals. However, for more complicated integrands---especially those arising from multifold MB integrals---this iterative integration may not be feasible to express in terms of MPLs, reflecting the underlying complexity of the problem. Even when the final result can be written in terms of MPLs, in most cases, it requires a significant amount of non-trivial manipulation. In the next section~\ref{sec:toGPL}, we demonstrate how the final results can be expressed in terms of Goncharov polylogarithms (GPLs) or iterated integrals, whenever such a representation is attainable.

\section{From real integrations to GPLs}
\label{sec:toGPL}
After reducing the MB integrals to integrals over real parameters, the resulting integrands typically consist of rational functions and logarithms of polynomials. This structure naturally suggests that the integrals can be expressed as iterated integrals, which, under favourable conditions, evaluate to GPLs. However, the path from these real integrations to explicit GPL representations is often highly nontrivial due to the increasing complexity of the integrands.
 
Once GPLs emerge in the intermediate expressions, a key step is to manipulate the integration variables such that they appear exclusively as the rightmost argument of the GPLs. This rearrangement is crucial because it enables the remaining integrals to be carried out systematically, ensuring that the final results are expressible purely in terms of GPLs. In simpler scenarios, where the GPL arguments are linear in the integration variables, this step can be efficiently performed using the fibration basis tools implemented in the {\tt PolyLogTools} package~\cite{Duhr:2019tlz}. The fibration basis method decomposes GPLs according to their arguments, facilitating variable shifts and integral evaluations in a controlled manner.

However, in many realistic cases, the integration variables enter the GPL weights as non-linear rational functions, substantially complicating their manipulation. Such non-linear dependencies obstruct straightforward application of the fibration basis and require more sophisticated techniques to isolate the integration variable at the rightmost position in each GPL.

Our strategy to overcome this involves exploiting the integral definition of GPLs to express a GPL with integration-variable-dependent weights in terms of lower-weight GPLs. Specifically, for a GPL of weight $n$ with argument-dependent weights, $G(\vec{a}_n(z); 1)$, where $\vec{a}_n(z) = \{a_1(z), a_2(z), \ldots, a_n(z)\}$ is an $n$-tuple of rational functions of the integration variable $z$~\cite{Duhr:2014woa}, we use the integral representation to write
\begin{align*}
    G(\vec{a}_n(z); 1) &= G(\vec{a}_n(z'); 1) \\
    &\quad + \int_{z'}^z dz\, \underbrace{\int_0^1 dt_1\, \frac{\partial}{\partial z} \left[ \frac{1}{t_1 - a_1(z)} G(\vec{a}_{n-1}(z); t_1) \right]}_{I_1}.
\end{align*}
Here, the boundary point $z'$ is chosen such that the GPL is regular at $z=z'$, ensuring the integrals are well-defined. The integrand term $I_1$ decomposes into a linear combination of rational functions multiplied by GPLs of weight $n-1$. By iterating this procedure, the original GPL is expressed in terms of GPLs with fewer and simpler weight arguments, each depending less intricately on the integration variable. We iterate this process until it reaches a form where we can shift the integration variable to the rightmost argument. For instance, if a resulting GPL has linear weights, the fibration basis can be used to shift the variable $z$ to the rightmost argument. 
Else, we manipulate the argument to isolate $z$ at the rightmost position inside each GPL, effectively rewriting
\[
G(\vec{a}_n(z); 1) \quad \to \quad G(\vec{b}_n; z),
\]
where the new weights $\vec{b}_n$ are independent of $z$. This form is crucial as it allows the subsequent integral over $z$ to be performed straightforwardly. Throughout this restructuring, it is essential to carefully regulate spurious singularities that appear at intermediate stages of the transformation to ensure that the final expressions remain finite and physically meaningful.

To implement this method systematically, we have developed dedicated in-house algorithms that automate the manipulation and reduction of GPLs with complex weight structures. This automation enables us to handle even highly nontrivial cases and express the final analytic results entirely in terms of GPLs. We verify the correctness and precision of our expressions through extensive numerical checks.
In the next section~\ref{sec:4den-massless}, we illustrate how, by following this approach, we successfully solve the case of four-denominator angular phase-space integrals in terms of GPLs.

\section{Four-denominator: massless}
\label{sec:4den-massless}
The four-denominator massless phase-space angular integral exhibits the following MB representation:
\begin{align}
 &\Omega_{1,1,1,1}(\{v_{ij}\},\epsilon)=\frac{1}{\Gamma (-2 \epsilon -2)} 2^{-2 \epsilon -2} \pi ^{1-\epsilon }\nonumber\\
 &\times \int_{-\infty}^{+\infty}\frac{dz_{12}dz_{13}dz_{14}dz_{23}dz_{24}dz_{34}}{(2\pi i)^6} v_{12}^{z_{12}} v_{13}^{z_{13}}
   v_{14}^{z_{14}} v_{23}^{z_{23}} v_{24}^{z_{24}} v_{34}^{z_{34}} \Gamma
   \left(-z_{12}\right) \Gamma \left(-z_{13}\right) \Gamma \left(-z_{14}\right) \nonumber\\
   &\times\Gamma
   \left(z_{12}+z_{13}+z_{14}+1\right) \Gamma \left(-z_{23}\right) \Gamma
   \left(-z_{24}\right) \Gamma \left(z_{12}+z_{23}+z_{24}+1\right) \Gamma
   \left(-z_{34}\right)\nonumber\\
   &\times\Gamma \left(z_{13}+z_{23}+z_{34}+1\right) \Gamma
   \left(z_{14}+z_{24}+z_{34}+1\right)\nonumber\\
   &\times\Gamma \left(-\epsilon
   -z_{12}-z_{13}-z_{14}-z_{23}-z_{24}-z_{34}-3\right)  \,.
\end{align}
For ease of presentation, we adopt a normalization factor and perform the calculation of
\begin{align}
\label{eq:MB-3denom-norm}
I^{(0)}_{1,1,1,1}&=C_\epsilon\Omega_{1,1,1,1}(\{v_{ij}\},\epsilon)
\end{align}
with 
\[
C_\epsilon=2^{-1+2\epsilon}\pi^{\epsilon}\frac{\Gamma(1-2\epsilon)}{\Gamma(1-\epsilon)}.
\] 
The superscript $(0)$ implies that the integral is massless. We expand $I^{(0)}_{1,1,1,1}$ in Laurent series in $\epsilon$ as 
\begin{align}
    I^{(0)}_{1,1,1,1} = \sum_{m=-1}^\infty \epsilon^m I^{(0),m}_{1,1,1,1}.
\end{align}
where the second superscript $``m"$ in $I^{(0),m}_{1,1,1,1}$ denotes the order of $\epsilon$. Following the steps discussed in section~\ref{sec:MB} and \ref{sec:ana-int}, we finally get a list of 10 integrations over real variables, which read as 
\begin{align}
\label{eq:real-int-massless}
I_1=&\int dv_1  dx_1\frac{\pi  v_{14} x_1}{8 v_{13} v_{34} \left(v_{12}^2 v_{34} \bar{x}_1-v_1 v_{14}
   v_{23} x_1\right)},\nonumber\\[6pt]
I_2=&\int dv_1 dx_1 \frac{-\pi  v_{23} v_{34} x_1}{8 v_{12} v_{13} \left(v_1 v_{34} \bar{x}_1-v_{14} v_{23}
   x_1\right)^2},\nonumber\\[6pt]
I_3=&\int dv_1 dx_1 dx_2 dx_3 \frac{3 \pi  v_1 v_{12}^5 v_{13}^3 v_{14} v_{23}^2 v_{24}^2 v_{34}^2 x_1
   x_3^2 \bar{x}_1 \bar{x}_2^2}{4 \left(v_1 v_{14} v_{23} \left(v_{12} v_{34}
   \bar{x}_1 \bar{x}_3+v_{13} v_{24} x_1 x_2\right)-v_{12}^2 v_{13} v_{24} v_{34} x_3
   \bar{x}_2\right)^4},\nonumber\\[6pt]
I_4=&\int dx_1 dx_2 dx_3 \frac{-\pi  v_{12}^2 v_{13}^2 v_{14} v_{23} v_{24} v_{34}^2 \bar{x}_1 \bar{x}_2
   \bar{x}_3}{2 \left(v_{12} v_{34} \left(v_{13} v_{24} \bar{x}_2 \bar{x}_3-v_{14}
   v_{23} x_2 \bar{x}_1\right)+v_{13} v_{14} v_{23} v_{24} x_1 x_3\right)^3},\nonumber\\[6pt]
I_5=&\int dv_1 dx_1 dx_2 dx_3 \frac{-3 \pi  v_1^2 v_{12}^3 v_{13}^2 v_{14}^3 v_{23}^2 v_{24}^3 v_{34}
   x_1^2 x_2 x_3 \bar{x}_2 \bar{x}_3}{4 \left(v_1 v_{13} v_{24} \left(v_{12} v_{34}
   \bar{x}_1 \bar{x}_3+v_{14} v_{23} x_1 x_2\right)-v_{12}^2 v_{14} v_{23} v_{34} x_3
   \bar{x}_2\right)^4},\nonumber\\[6pt]
I_6=& \int dx_1 dx_2 dx_3 \frac{\pi  v_{12}^2 v_{13} v_{14}^2 v_{23} v_{24} v_{34}^2 x_3 \bar{x}_1
   \bar{x}_2}{2 \left(v_{12} v_{34} \left(v_{14} v_{23} \bar{x}_1 \bar{x}_2+v_{13}
   v_{24} x_2 x_3\right)-v_{13} v_{14} v_{23} v_{24} x_1 \bar{x}_3\right)^3},\nonumber\\[6pt]
I_7=&\int dv_1 dx_1 dx_2 dx_3\frac{3 \pi  v_1^2 v_{12}^3 v_{13}^2 v_{14}^2 v_{23}^2 v_{24}^3 v_{34}^2
   x_1 x_2 x_3 \bar{x}_1 \bar{x}_2 \bar{x}_3}{2 \left(v_{14} v_{23} v_{34} v_{12}^2
   x_3 \bar{x}_1+v_1 v_{13} v_{24} \left(v_{14} v_{23} x_1 \bar{x}_2+v_{12} v_{34} x_2
   \bar{x}_3\right)\right)^4},\nonumber\\[6pt]
I_8=&\int dv_1 dx_1 dx_2 dx_3 \frac{-3 \pi  v_1^3 v_2^2 v_{12} v_{13}^2 v_{14}^2 v_{23}^3 v_{24}^3
   v_{34}^4 x_1 x_2 x_3^2 \bar{x}_1^2 \bar{x}_2^2 \bar{x}_3}{\left(v_1 v_{34}
   \left(v_{12} v_{14} v_{23} \bar{x}_1 \bar{x}_2+v_2 v_{13} v_{24} x_2 x_3\right)-v_2
   v_{13} v_{14} v_{23} v_{24} x_1 \bar{x}_3\right)^5},\nonumber\\[6pt]
I_9=&\int dx_1 \frac{\pi }{-4 v_{12} v_{13} v_{34} x_1+4 v_{13} v_{14} v_{23} x_1+4 v_{12} v_{13}
   v_{34}},\nonumber\\[6pt]
I_{10}=&\int dx_1 dx_2 dx_3 \frac{-\pi  v_{12} v_{13} v_{14} v_{23} v_{24} v_{34}}{4 \left(v_{12} v_{14} v_{23}
   v_{34} x_3 \bar{x}_1+v_{13} v_{24} \left(v_{14} v_{23} x_1 \bar{x}_2+v_{12} v_{34}
   x_2 \bar{x}_3\right)\right)^2}   \,,
\end{align}
where we define $\bar{x}_i = 1 - x_i$. The integration ranges are $v_1 \in [0, v_{12}]$ and $x_i \in [0,1]$.
Utilising the method discussed in section~\ref{sec:toGPL}, we solve the integrals and express the final result of $I^{(0)}_{1,1,1,1}$ in terms of GPLs. For this massless case, we obtain weight 2 GPLs with the following 17 letters:
\begin{align}
l_1   & = 0, & l_2   & = 1, & l_3   & = \frac{v_{13} v_{24}}{X_4}, & l_4   & = \frac{v_{14} v_{23}}{X_5}, \nonumber \\
l_5   & = \frac{v_{13} v_{24}}{X_6}, & l_6   & = -\frac{v_{12} v_{34}}{X_5}, & l_7   & = -\frac{v_{12} v_{34}}{X_6}, & l_{8,13}   & = \frac{X_1 \mp \sqrt{Y_1}}{2 v_{14} v_{23}}, \nonumber \\
l_{9,14}   & = \frac{X_1 \mp \sqrt{Y_1}}{2 v_{13} v_{24}}, & l_{10,15} & = \frac{X_2 \mp \sqrt{Y_1}}{2 v_{14} v_{23}}, & l_{11,16} & = \frac{X_3 \mp \sqrt{Y_1}}{2 v_{13} v_{24}}, & l_{12,17} & = \frac{X_3 \mp \sqrt{Y_1}}{2 v_{12} v_{34}}
\end{align}
with
\begin{align}
X_1 &= v_{14} v_{23} + v_{13} v_{24} - v_{12} v_{34}, & X_2 &= v_{14} v_{23} - v_{13} v_{24} + v_{12} v_{34}, \nonumber \\
X_3 &= -v_{14} v_{23} + v_{13} v_{24} + v_{12} v_{34}, & X_4 &= -v_{14} v_{23} + v_{13} v_{24}, \nonumber \\
X_5 &= v_{14} v_{23} - v_{12} v_{34}, & X_6 &= v_{13} v_{24} - v_{12} v_{34},\nonumber\\
Y_1 &= v_{14}^2 v_{23}^2 - 2 v_{14} v_{23} (v_{13} v_{24} + v_{12} v_{34}) + X_6^2. & &
\end{align}
We present the analytic expression for $I^{(0)}_{1,1,1,1}$ in the ancillary file {\tt <massless_ep0.m>} up to $\mathcal{O}(\epsilon^0)$, i.e., $I^{(0),0}_{1,1,1,1}$, in terms of GPLs.
Notably, within our computational approach, no square roots arise during the evaluation of the integrals themselves; the quantities 
$Y_i$ appear only as square roots in the letters once the iterated 
integrals are rewritten in terms of GPLs. The massless angular 
integral features a single $\epsilon^{-1}$ pole originating from a 
collinear singularity, which turns out to be
%
\begin{align}
I^{(0),-1}_{1,1,1,1} =  -\frac{\pi}{8} \frac{1}{4!} \sum_{\sigma \in S_4} \frac{v_{\sigma(1)\sigma(2)} v_{\sigma(1)\sigma(3)} v_{\sigma(2)\sigma(3)}}{\prod_{1 \le i < j \le 4} v_{ij}}.
\end{align}
This matches with the result given in ref.~\cite{Smirnov:2024pbj}. The expression is explicitly symmetrized over the symmetry group $S_4$, reflecting the expected invariance under permutations of the indices $\{1,2,3,4\}$. 
We numerically verify the permutation symmetry of this finite term, providing a strong consistency check of the result. Additionally, our numerical cross-checks of the angular integral are performed using \texttt{GiNaC} through \texttt{PolyLogTools} framework and the package {\tt MB.m}~\cite{Czakon:2005rk}. We find an agreement, as shown in table~\ref{tab:massless}.
\begin{table}[h!]
\centering
\begin{tabular}{|c|c|c|c|}
\hline
$I^{(0)}_{1,1,1,1}$ & Set 1 & Set 2 & Set 3 \\ \hline
Our results & $ \frac{-30.7399}{\epsilon}-21.3574$ &  $\frac{ -286.299}{\epsilon}-566.3291$ & $\frac{-92.3456}{\epsilon}-103.21584$\\ \hline
Direct evaluation & $\frac{-30.7399}{\epsilon}-21.3574$ &$\frac{ -286.299}{\epsilon} -566.3291$ & $\frac{-92.3456}{\epsilon}-103.216$ \\ \hline
\end{tabular}
\caption{Comparison of our results against direct numerical evaluation using the MB program. Three sets of phase-space points are $\{v_{12}, v_{13}, v_{14}, v_{23}, v_{24}, v_{34}\}: $ \{0.45, 0.40, 0.35, 0.30, 0.35, 0.40\}, \{0.10, 0.15, 0.30, 0.25, 0.115, 0.35\}, \{0.32, 0.23, 0.16, 0.28, 0.37, 0.27\}. }
\label{tab:massless}
\end{table}
While the MB program takes about half an hour to evaluate the angular integral at one phase-space point, our result requires only about a second. In the following section~\ref{sec:masive}, we extend the discussion to the massive case. 

\section{Four-denominator: massive}
\label{sec:masive}
The vectors, $p_i^\mu$, appearing in the definition of the angular integral in \eqref{eq:ang_int} can also be massive. In the following subsections, we discuss various scenarios where one or more of these vectors carry mass.
\subsection{Single-massive}
The four denominator single-mass phase space angular integral $\Omega_{1,1,1,1}$ admits the following MB representation:
\begin{align}
\Omega_{1,1,1,1}&(v_{11},\{v_{ij}\},\epsilon)= 
\frac{1}{\Gamma (-2 \epsilon -2)} 2^{-2 \epsilon -2} \pi ^{1-\epsilon }\nonumber\\
 &\times \int_{-\infty}^{+\infty}\frac{dz_{11}dz_{12}dz_{13}dz_{14}dz_{23}dz_{24}dz_{34}}{(2\pi i)^7} v_{11}^{z_{11}}v_{12}^{z_{12}} v_{13}^{z_{13}}
   v_{14}^{z_{14}} v_{23}^{z_{23}} v_{24}^{z_{24}} v_{34}^{z_{34}}\Gamma\left(-z_{11}\right) \Gamma
   \left(-z_{12}\right)\nonumber\\
   &\times\Gamma \left(-z_{13}\right) \Gamma \left(-z_{14}\right) 
   \Gamma
   \left(2z_{11}+z_{12}+z_{13}+z_{14}+1\right) \Gamma \left(-z_{23}\right) \Gamma
   \left(-z_{24}\right) \nonumber\\
   &\times\Gamma \left(z_{12}+z_{23}+z_{24}+1\right) \Gamma
   \left(-z_{34}\right)\nonumber\\
   &\times\Gamma \left(z_{13}+z_{23}+z_{34}+1\right) \Gamma
   \left(z_{14}+z_{24}+z_{34}+1\right)\nonumber\\
   &\times\Gamma \left(-\epsilon-z_{11}
   -z_{12}-z_{13}-z_{14}-z_{23}-z_{24}-z_{34}-3\right)  \,.
\end{align}
Without loss of generality, we assume $p_1^\mu$ to be massive. Similar to the massless case, we include the same normalisation factor, $C_\epsilon$, and compute 
\begin{align}
\label{eq:MB-4denom-norm}
I^{(1)}_{1,1,1,1}&=C_\epsilon\Omega_{1,1,1,1}(v_{11},\{v_{ij}\},\epsilon).
\end{align}
The superscript $(1)$ indicates that the integral has a single mass. Similar to the massless case, we expand $I^{(1)}_{1,1,1,1}$ in Laurent series in $\epsilon$ as 
\begin{align}
    I^{(1)}_{1,1,1,1} = \sum_{m=-1}^\infty \epsilon^m I^{(1),m}_{1,1,1,1}.
\end{align}
Following the procedure outlined in sections~\ref{sec:MB} and~\ref{sec:ana-int}, we obtain a set of 14 new integrals over real variables of the form
\begin{align}
I_i = \int d\mu_i \frac{A_i}{B_i},
\end{align}
where the integration measures $d\mu_i$ and the rational functions $A_i/B_i$ depend on the kinematic variables and parameters. The integration variables are $\{x_i, v_1\}$, with $v_1 \in [0, v_{12}]$ and $x_i \in [0, 1]$. Depending on the variables appearing in the integrand, $d\mu_i$ is defined over the corresponding subset. The explicit forms of $A_i$ and $B_i$ for $i=1,\dots,15$ are presented below:
\begin{align}
A_1 &= \pi v_{13} x_1, \nonumber\\
B_1 &= -8 v_{11} v_{12} v_{24} v_{23}^2 x_1 
      + 8 v_{1} v_{12} v_{13} v_{24} v_{23} x_1
      + 8 v_{11} v_{12} v_{24} v_{23}^2, \nonumber\\
A_2 &= \pi v_{1} v_{14}^2 x_2^2, \nonumber\\
B_2 &= 8 v_{34} \left( v_{11} v_{12} v_{34} (x_2 - 1) - v_{1} v_{13} v_{14} x_2 \right)
       \left( v_{12}^2 v_{34} (x_2 - 1) - v_{1} v_{14} v_{23} x_2 \right), \nonumber\\[1ex]
A_3 &= \pi v_{14} x_2, \nonumber\\
B_3 &= 4 \left( v_{11} v_{34} (x_2 - 1) - v_{13} v_{14} x_2 \right)
       \left( v_{12} v_{34} (x_2 - 1) - v_{14} v_{23} x_2 \right), \nonumber\\[1ex]
A_4 &= \pi v_{34} (x_2 - 1) x_2,\nonumber \\
B_4 &= 8 \left( v_{1} v_{11} v_{34} (x_2 - 1) - v_{12} v_{13} v_{14} x_2 \right)
       \left( v_{1} v_{34} (x_2 - 1) - v_{14} v_{23} x_2 \right),\nonumber\\
A_5 &= -\pi v_{11}^2 v_{12} v_{14}^3 v_{23}^2 v_{24} \bar{x}_1^2 x_3 \bar{x}_4, \nonumber\\
B_5 &= 2(v_{12} v_{14} x_1 (v_{14} v_{23} x_3 + v_{13} v_{24} \bar{x}_1) \bar{x}_3 x_4) + v_{11} v_{24} \bar{x}_1 (v_{14} v_{23} \bar{x}_4 - v_{12} v_{34} \bar{x}_1 \bar{x}_3 x_4))^3, \nonumber\\
A_6 &= \pi v_{11}^2 v_{12} v_{13} v_{14} v_{34}^2 \bar{x}_1^2 x_2 x_4 ,\nonumber\\
B_6 &= 2v_{23} (v_{11} v_{34} \bar{x}_2 \bar{x}_1^2 (v_{12} v_{34} x_4 - v_{13} v_{24} x_2 \bar{x}_4) + v_{13} v_{14} x_1 x_2 (v_{14} v_{23} x_2 \bar{x}_4 - v_{12} v_{34} \bar{x}_2 \bar{x}_1))^3 ,\nonumber\\
A_7 &= 3\pi v_{11}^5 v_{11}^3 v_{13}^2 v_{14} v_{34}^6 \bar{x}_2^2 x_4 \bar{x}_1^6 \bar{x}_2^4 x_4, \nonumber\\
B_7 &= 4v_{23} (v_{12} v_{13} v_{14}^2 v_{23} x_1 x_2 \bar{x}_4 - v_{11} v_{13} v_{34} x_2 \bar{x}_1 \bar{x}_2 (v_{11} v_{24} \bar{x}_1 \bar{x}_4 + v_{12} v_{14} x_1) \nonumber\\
&+ v_{11} v_{34} \bar{x}_2 \bar{x}_1 \bar{x}_2)^4 ,\nonumber\\
A_8 &= \pi v_{11} v_{12} v_{13}^2 v_{14}^2 v_{23} v_{34}^2 \bar{x}_1^3 x_2^3 \bar{x}_2 \bar{x}_4, \nonumber\\
B_8 &= 2 (v_{11} v_{34} \bar{x}_2 \bar{x}_1^2 (v_{13} v_{24} x_2 \bar{x}_4 - v_{12} v_{34} x_4) + v_{13} v_{14} x_1 x_2 (v_{12} v_{34} \bar{x}_1 \bar{x}_2 - v_{14} v_{23} x_2 \bar{x}_4))^3, \nonumber\\
A_9 &= 3\pi v_{11}^2 v_{11}^3 v_{12}^2 v_{14}^2 v_{23}^2 v_{24}^3 x_1 x_3 x_4 \bar{x}_1^3 \bar{x}_3 \bar{x}_4, \nonumber\\
B_9 &= 2 (v_{11} v_{24} \bar{x}_1 (v_{11} v_{12} v_{34} x_4 (\bar{x}_1 - x_1 x_3 + x_3) + v_{12} v_{13} v_{14} x_1 x_4 \bar{x}_3 + v_{11} v_{14} v_{23} \bar{x}_4) \nonumber\\
&+ v_{12} v_{23} v_{14}^2 x_1 x_3)^4 ,\nonumber\\
A_{10} &= 3\pi v_{11}^2 v_{12} v_{14}^2 v_{23}^2 v_{24}^2 v_{34} x_1 x_3 x_4 \bar{x}_1^3 \bar{x}_3 \bar{x}_4, \nonumber\\
B_{10} &= 2(v_{12} v_{14} x_1 (v_{13} v_{24} x_4 \bar{x}_1 \bar{x}_3 + v_{14} v_{23} x_3) + v_{11} v_{24} \bar{x}_1 (v_{11} v_{34} x_4 (\bar{x}_1 - x_1 x_3 + x_3) \nonumber\\
&+ v_{14} v_{23} \bar{x}_4))^4, \nonumber\\
A_{11} &= 3\pi v_{11}^4 v_{11}^2 v_{12}^6 v_{13}^3 v_{14}^3 v_{23} v_{24} v_{34}^5 x_1^2 x_2^5 x_4 \bar{x}_1^6 \bar{x}_2^3 \bar{x}_4^2 ,\nonumber\\
B_{11} &= (v_{12}^2 v_{13} v_{14}^2 v_{23} x_1 x_2 \bar{x}_4 - v_{11} v_{13} v_{34} x_2 \bar{x}_1 \bar{x}_2 (v_{12} v_{11} v_{24} \bar{x}_1 \bar{x}_4 + v_{14} v_{12}^2 x_1) \nonumber\\
&+ v_{11} v_{12} v_{34} \bar{x}_2 \bar{x}_1 \bar{x}_2)^5, \nonumber\\
A_{12} &= -3\pi v_{11}^2 v_{11}^3 v_{12}^3 v_{14}^3 v_{23}^2 v_{24}^2 \bar{x}_1^4 \bar{x}_3 x_3 \bar{x}_4 x_4, \nonumber\\
B_{12} &= 4 (v_{11} v_{14} v_{23} (v_{12} v_{14} x_1 x_3 + v_{11} v_{24} \bar{x}_1 \bar{x}_4) - v_{12}^2 v_{24} \bar{x}_1 \bar{x}_3 x_4 (v_{11} v_{34} \bar{x}_1 - v_{13} v_{14} x_1))^4, \nonumber\\
A_{13} &= -\pi v_{11}^2 v_{12} v_{14}^3 v_{23}^2 v_{24}^2 \bar{x}_1^2 x_3 \bar{x}_4, \nonumber\\
B_{13} &= 2(v_{12} v_{14} x_1 (v_{14} v_{23} x_3 + v_{13} v_{24} \bar{x}_1) \bar{x}_3 x_4) + v_{11} v_{24} \bar{x}_1 (v_{14} v_{23} \bar{x}_4 - v_{12} v_{34} \bar{x}_1 \bar{x}_3 x_4))^3,\nonumber\\
A_{14} &= \pi, \nonumber\\
B_{14} &= {-8 v_{11} v_{14} v_{23} x_1+8 v_{12} v_{13} v_{14} x_1+8 v_{11} v_{14} v_{23}},\nonumber\\
A_{15} &= \pi  v_{11} v_{12} v_{14}{}^2 v_{23} v_{24} \bar{x}_1, \nonumber\\
B_{15} &= 4 \Big(v_{12} v_{14} x_1 (v_{14} v_{23} x_3+v_{13} v_{24} \bar{x}_1 \bar{x}_3 x_4)+v_{11} v_{24} \bar{x}_1 (v_{14} v_{23} \bar{x}_4+v_{12} v_{34} \bar{x}_1 \bar{x}_3
   x_4)\Big)^2\,.
\end{align}
In comparison to the massless case, these integrals are more complicated due to the involvement of an additional variable $v_{11}$. By employing the methodology discussed in section~\ref{sec:toGPL}, we express these integrals in terms of GPLs. The highest transcendental weight of the appearing GPLs is 2, similar to the massless case. At ${\cal O}(\epsilon^0)$, we obtain the following set of 11 letters in $I^{(1),0}_{1,1,1,1}$:
\begin{align}
l_1^{(1)} &=\frac{v_{11}v_{23}}{v_{12}v_{13}}, &
l_2^{(1)} &=\frac{v_{11}v_{23}}{-v_{12}v_{13} + v_{11}v_{23}},  &
l_3^{(1)} &=\frac{v_{11}v_{24}}{-v_{12}v_{14} + v_{11}v_{24}}, \nonumber\\
l_4^{(1)} &=\frac{v_{13}v_{24}}{-v_{14}v_{23} + v_{13}v_{24}}, &
l_5^{(1)} &=\frac{v_{11}v_{14}v_{23} - v_{11}v_{13}v_{24}}{v_{12}v_{13}v_{14} - v_{11}v_{13}v_{24}}, &
l_6^{(1)} &=\frac{v_{11}v_{34}}{-v_{13}v_{14} + v_{11}v_{34}}, \nonumber\\
l_7^{(1)} &=\frac{v_{12}v_{34}}{-v_{14}v_{23} + v_{12}v_{34}}, &
l_8^{(1)} &=\frac{v_{12}v_{34}}{-v_{13}v_{24} + v_{12}v_{34}}, &
l_9^{(1)} &= \frac{v_{11}v_{14}v_{23} - v_{11}v_{12}v_{34}}{v_{12}v_{13}v_{14} - v_{11}v_{12}v_{34}}, \nonumber
\end{align}
\begin{align}
    l_{10,11}^{(1)} =\frac{Z_{1,2}^{(1)}}{2 (v_{12} v_{14} - v_{11} v_{24})(v_{13} v_{14} - v_{11} v_{34})}.    
\end{align}
In the aforementioned expressions, we define
\begin{align}
Z_{1,2}^{(1)}&=v_{11} \Big( 
   v_{14}^2 v_{23} + 2 v_{11} v_{24} v_{34} + 
   v_{14} \big(
     -v_{13} v_{24} - v_{12} v_{34} \nonumber\\ &\pm
     \sqrt{(v_{14} v_{23} - v_{13} v_{24})^2 
       - 2 (v_{12} v_{14} v_{23} + v_{12} v_{13} v_{24} - 2 v_{11} v_{23} v_{24}) v_{34} 
       + v_{12}^2 v_{34}^2}
   \big) 
\Big).
\end{align}
We present the analytic expression for $I^{(1),0}_{1,1,1,1}$ in the ancillary file {\tt <one_mass_ep0.m>} up to $\mathcal{O}(\epsilon^0)$ in terms of GPLs. The file is {\tt Mathematica} readable.
Notably, within our computational approach, no square roots arise during the evaluation of the integrals themselves; the square roots inside 
$X^{(1)}_{1,2}$ appear only after the iterated 
integrals are rewritten in terms of GPLs. This is similar to the massless case. The single-massive angular 
integral features a $\epsilon^{-1}$ pole originating from a 
collinear singularity, which turns out to be
\begin{align}
I^{(1),-1}_{1,1,1,1} = 
-\frac{\pi}{8} \frac{1}{2} 
\frac{
\displaystyle  \sum_{\sigma \in S_3} v_{1\,\sigma(2)} v_{1\,\sigma(3)} v_{\sigma(2)\,\sigma(3)}
}{
\displaystyle \prod_{1 \le i < j \le 4} v_{ij}
}.
\end{align}
Here, $S_3$ denotes the symmetric group consisting of all permutations of the set $\{2,3,4\}$. The pole agree with the result given in ref.~\cite{Smirnov:2024pbj,Salvatori:2024nva}. The permutation symmetry of the finite term is verified numerically, providing a strong consistency check on our result. We verify the finite term numerically against direct numerical evaluation of the MB representation of angular integral using MB program and we find a perfect agreement, as shown in table~\ref{tab:massive}. 
\begin{table}[h!]
\centering
\begin{tabular}{|c|c|c|c|}
\hline
$I^{(1)}_{1,1,1,1}$ & Set 1 & Set 2 & Set 3 \\ \hline
Our results &$\frac{-24.5066}{\epsilon} -16.4429 $ & $\frac{-199.033}{\epsilon}-520.0359 $&  $\frac{ -58.9982}{\epsilon}-88.0371 $\\ \hline
Direct evaluation &$\frac{-24.5066}{\epsilon} -16.5783 $ & $\frac{-199.033}{\epsilon}-520.0358$& $\frac{ -58.9982 }{\epsilon}-88.0371$\\ \hline
\end{tabular}
\caption{Comparison of our results against direct numerical evaluation using the MB program. Three sets of phase-space points are $\{v_{11}, v_{12}, v_{13}, v_{14}, v_{23}, v_{24}, v_{34}\}: $ \{0.11, 0.45, 0.40, 0.35, 0.30, 0.35, 0.40\}, \{0.15, 0.10, 0.15, 0.30, 0.25, 0.115, 0.35\}, \{0.2, 0.32, 0.23, 0.16, 0.28, 0.37, 0.27\}. }
\label{tab:massive}
\end{table}
Whereas the MB program takes around half an hour to evaluate the angular integral at a single phase-space point, our result can be obtained in only about one second. Our results also numerically agree with ref.~\cite{Salvatori:2024nva}.

\subsection{Double-,triple-, and quartic-massive}
\label{sec:double-massive}
The methodology discussed in the previous sections can be applied to obtain the result for the double-massive angular integral. However, it turns out that one can avoid performing an explicit computation of this integral. Instead, by using partial fraction decomposition~\cite{Lyubovitskij:2021ges}, this integral can be related to the single-massive case. This is possible because angular integrals are linear in parameter space when expressed in a suitable coordinate system, allowing such a decomposition.
In the cited reference, the two-point splitting lemma is used to decompose fully massive two-denominator angular integrals into linear combinations of single-massive ones. Here, we extend this approach to a four-denominator double-massive angular integral by partially fractioning it with respect to its two massive denominators, leading to
 \begin{align}
 \label{eq:pf-double}
 &{I}^{(2)}_{1,1,1,1}
\left(v_{11},v_{22},v_{12},v_{13},v_{14},v_{23},v_{24},v_{34}\},\epsilon\right)\nonumber\\
=   &\lambda_{12} {I}^{(1)}_{1,1,1,1}\left(v_{11},v_{13},v_{14},v_{15},
   v_{34},v_{35},v_{45}\right) \nonumber\\
   &+\bar\lambda_{12} {I}^{(1)}_{1,1,1,1}\left(v_{22},v_{23},v_{24},v_{25},
   v_{34},v_{35},v_{45}\right),
   \end{align}
where
\begin{align}
&\lambda_{ij} = \frac{2v_{ii}-v_{ij}-\sqrt{v_{ij}^2-4v_{ii}v_{jj}}}{2v_{ii}-2v_{ij}+2v_{jj}}, \bar\lambda_{ij}=1-\lambda_{ij}, \quad {\rm and} \quad v_{ij}=v_{ji}\,.
\end{align}
The last identity follows directly from the definition~\eqref{eq:sp}. The parameter $\lambda_{ij}$ is a function of the scalar products $v_{ij}$.
The decomposition introduces new scalar products $v_{i5}$, which can be expressed in terms of old ones. Their explicit forms are defined in \eqref{eq:new-sp}. 
Without any loss of generality, in our analysis, we assume the vectors $p_1^\mu$ and $p_2^\mu$ to be massive.

By repeatedly applying partial fraction decomposition, we can express the four denominator triple-massive angular integral in terms of single-massive ones. We get
\begin{align}
 \label{eq:pf-triple}
&{I}^{(3)}_{1,1,1,1}\left(v_{11},v_{22},v_{33},v_{12},v_{13},v_{14},v_{23},v_{24},v_{34}\right)\nonumber\\
=& \lambda_{12}\lambda_{13}  {I}^{(1)}_{1,1,1,1}\left(v_{11},v_{14},v_{15},v_{1
   6},v_{45},v_{46},v_{56}\right)\nonumber\\
+ &\lambda_{12}\bar\lambda_{13}  {I}^{(1)}_{1,1,1,1}\left(v_{33},v_{34},v_{35},v_{3
   6},v_{45},v_{46},v_{56}\right)\nonumber\\
+ &\bar\lambda_{12}\lambda_{13}  {I}^{(1)}_{1,1,1,1}\left(v_{22},v_{24},v_{25},v_{27},v_{45},v_{47},v_{57}\right)\nonumber\\
+ &\bar\lambda_{12}\bar\lambda_{13}  {I}^{(1)}_{1,1,1,1}\left(v_{33},v_{34},v_{35},v_{3
   7},v_{45},v_{47},v_{57}\right)
\end{align}
Following the same partial fraction decomposition procedure, we obtain the result for the fully massive four-denominator phase-space angular integral in terms of single-massive integrals, which reads as
\begin{align}
 \label{eq:pf-quadrupole}
&{I}^{(4)}_{1,1,1,1}\left(v_{11},v_{22},v_{33},v_{44},v_{12},v_{13}
   ,v_{14},v_{23},v_{24},v_{34}\right)\nonumber\\
=   &\bar\lambda_{12}\lambda_{23}\bar\lambda_{24}{I}^{(1)}_{1,1,1,1}\left(v_{44},v_{45},v_{4
   7},v_{410},v_{57},v_{510},v_{710}\right)\nonumber\\
   &+\lambda_{12}\lambda_{13}\lambda_{14}{I}^{(1)}_{1,1,1,1}\left(v_{11},v_{15},v_{1
   6},v_{18},v_{56},v_{58},v_{68}\right)\nonumber\\
   &+\bar\lambda_{12}\lambda_{23}\bar\lambda_{14}{I}^{(1)}_{1,1,1,1}\left(v_{44},v_{45},v_{4
   6},v_{48},v_{56},v_{58},v_{68}\right)\nonumber\\
   &+\bar\lambda_{12}\lambda_{23}\lambda_{24}{I}^{(1)}_{1,1,1,1}\left(v_{22},v_{25},v_{27},v_{210},v_{57},v_{510},v_{710}\right)\nonumber\\
   &+\bar\lambda_{12}\bar\lambda_{23}\bar\lambda_{34}{I}^{(1)}_{1,1,1,1}\left(v_{44},v_{45},v_{4
   7},v_{48},v_{57},v_{58},v_{78}\right)\nonumber\\
   &+\lambda_{12}\bar\lambda_{13}\bar\lambda_{34}{I}^{(1)}_{1,1,1,1}\left(v_{44},v_{45},v_{4
   6},v_{49},v_{56},v_{59},v_{69}\right)\nonumber\\
   &+\bar\lambda_{12}\lambda_{23}\lambda_{34}{I}^{(1)}_{1,1,1,1}\left(v_{33},v_{35},v_{
37},v_{38},v_{57},v_{58},v_{78}\right)\nonumber\\
   &+\lambda_{12}\bar\lambda_{13}\bar\lambda_{34}{I}^{(1)}_{1,1,1,1}\left(v_{33},v_{35},v_{3
   6},v_{39},v_{56},v_{59},v_{69}\right)\,,
\end{align}
where
\begin{align}
\label{eq:new-sp}
v_{15}&=\bar{\lambda}_{12} 2v_{11}+\lambda _{12} v_{12},
v_{25}=\bar{\lambda}_{12} v_{12}+\lambda _{12} 2v_{22},\nonumber\\
v_{35}&=\bar{\lambda}_{12} v_{13}+\lambda _{12} 2v_{33},
v_{45}=\bar{\lambda}_{12} v_{14}+\lambda _{12} 2v_{44},\nonumber\\
v_{16}&=\bar{\lambda}_{13} 2v_{11}+\lambda _{13} v_{13},
v_{26}=\bar{\lambda}_{13} v_{12}+\lambda _{13} v_{23},\nonumber\\
v_{36}&=\bar{\lambda}_{13} v_{13}+\lambda _{13} 2v_{33},
v_{46}=\bar{\lambda}_{13} v_{14}+\lambda _{13} 2v_{44},\nonumber\\
v_{56}&=\bar{\lambda}_{12}\bar{\lambda}_{13} 2v_{11}+\bar{\lambda}_{12}\lambda _{13} v_{13}+\lambda _{12}\bar{\lambda}_{13}v_{12}+\lambda _{12} \lambda _{13} v_{23},\nonumber\\
v_{27}&=\bar{\lambda}_{23} 2v_{22}+\lambda _{23} v_{23},
v_{37}=\bar{\lambda}_{23} v_{23}+\lambda _{23} 2v_{33},\nonumber\\
v_{47}&=\bar{\lambda}_{23} v_{24}+\lambda _{23} 2v_{44},
v_{57}=\bar{\lambda}_{12}\bar{\lambda}_{23} v_{13}+\bar{\lambda}_{12}\lambda _{23} v_{13}+\lambda _{12}\bar{\lambda}_{23}2v_{22}+\lambda _{12} \lambda _{23} v_{23},\nonumber\\
v_{18}&=\bar{\lambda}_{14} 2v_{11}+\lambda _{14} v_{14},
v_{38}=\bar{\lambda}_{14} v_{13}+\lambda _{14} v_{34},\nonumber\\
v_{48}&=\bar{\lambda}_{14} v_{14}+\lambda _{14} 2v_{44},\nonumber\\
v_{58}&=\bar{\lambda}_{12}\bar{\lambda}_{14} 2v_{11}+\bar{\lambda}_{12}\lambda _{14} v_{14}+\lambda _{12}\bar{\lambda}_{14}v_{12}+\lambda _{12} \lambda _{14} v_{24},\nonumber\\
v_{68}&=\bar{\lambda}_{13}\bar{\lambda}_{14} 2v_{11}+\bar{\lambda}_{13}\lambda _{14} v_{14}+\lambda _{13}\bar{\lambda}_{14}v_{13}+\lambda _{13} \lambda _{14} v_{34},\nonumber\\
v_{78}&=\bar{\lambda}_{23}\bar{\lambda}_{14} v_{12}+\bar{\lambda}_{23}\lambda _{14} v_{24}+\lambda _{23}\bar{\lambda}_{14}v_{13}+\lambda _{23} \lambda _{34} v_{23},\nonumber\\
v_{39}&=\bar{\lambda}_{34} 2v_{33}+\lambda _{34} v_{34},
v_{49}=\bar{\lambda}_{34} v_{34}+\lambda _{34} 2v_{44},\nonumber\\
v_{59}&=\bar{\lambda}_{12}\bar{\lambda}_{34} v_{13}+\bar{\lambda}_{12}\lambda _{34} v_{14}+\lambda _{12}\bar{\lambda}_{34}v_{23}+\lambda _{12} \lambda _{24} v_{23},\nonumber\\
v_{69}&=\bar{\lambda}_{13}\bar{\lambda}_{34} v_{13}+\bar{\lambda}_{13}\lambda _{34} v_{14}+\lambda _{13}\bar{\lambda}_{34}2v_{33}+\lambda _{13} \lambda _{34} v_{23},\nonumber\\
v_{210}&=\bar{\lambda}_{22} 2v_{22}+\lambda _{24} v_{24},
v_{410}=\bar{\lambda}_{24} v_{24}+\lambda _{24} 2v_{44},\nonumber\\
v_{510}&=\bar{\lambda}_{12}\bar{\lambda}_{24} v_{12}+\bar{\lambda}_{12}\lambda _{24} v_{14}+\lambda _{12}\bar{\lambda}_{24}2v_{22}+\lambda _{12} \lambda _{34} v_{24},\nonumber\\
v_{710}&=\bar{\lambda}_{23}\bar{\lambda}_{24} 2v_{22}+\bar{\lambda}_{23}\lambda _{24} v_{24}+\lambda _{23}\bar{\lambda}_{24}v_{23}+\lambda _{23} \lambda _{34} v_{34}.\nonumber
\end{align}

In the ancillary files, we provide explicit results for the massless $I^{(0),0}_{1,1,1,1}$ and the single-massive angular integral $I^{(1),0}_{1,1,1,1}$. For the multi-mass cases, we explicitly implement the partial-fraction relations~\eqref{eq:pf-double}, \eqref{eq:pf-triple}, and \eqref{eq:pf-quadrupole}, from which the results can be obtained straightforwardly. All expressions are given in terms of GPLs. The massless and single-massive results are further optimized for fast numerical evaluation, and in all cases the evaluation takes only about a second using {\tt GiNaC}, reflecting the high level of optimization. The numerical evaluation of our expressions can be found in {\tt <main.nb>}. Since the results are expressed directly in terms of GPLs---unlike in ref.~\cite{Haug:2025sre}---they can be readily combined with the radial components of a given scattering process, thanks to the iterative nature of GPLs. We numerically cross-check the massless and single-massive MB integrals against our resulting expression in terms of GPLs. Moreover, we explicitly confirm the expected permutation symmetries in the finite parts.

\section{Recursion relations}
\label{sec:recursion}
Angular integrals containing higher powers of propagator denominators are not independent objects: they can be systematically related to integrals with lower powers through a set of recursion relations. These relations follow from the differential identities
\begin{align}
    \frac{\partial}{\partial v_{kl}} \,
    I^{(n)}_{j_1,j_2,j_3,j_4} 
    = \sum_{i_1,i_2,i_3,i_4} 
      C_{kl}^{(i_1 i_2 i_3,i_4)} \,
      I^{(n)}_{i_1,i_2,i_3,i_4},
\end{align}
where $v_{kl}$ denotes the dimensionless scalar product of the reference 
momenta $p_k$ and $p_l$. Differentiating with respect to $v_{kl}$ probes the dependence of the angular integral on the underlying kinematic configuration, and in doing so, generates integrals with shifted denominator powers. The coefficients $C_{kl}^{(i_1 i_2 i_3,i_4)}$ are rational functions of the kinematic invariants $v_{ij}$, the dimensional regulator $\epsilon$, and the integer indices $i_1, i_2, i_3,i_4$.

By iteratively applying these equations, any angular integral with arbitrarily 
high denominator powers can be reduced to a finite basis of independent 
objects, commonly referred to as \emph{master integrals}. This reduction 
removes the need to evaluate each higher-power case from scratch: once the 
master integrals are known, all others follow by purely algebraic substitution. 
In this respect, the recursion relations play a role closely analogous to 
integration-by-parts (IBP) identities~\cite{Chetyrkin:1981qh,Laporta:2000dsw} 
in loop momentum space, but here the reduction takes place in the space of 
angular variables. This makes the method especially suitable for factorized 
phase-space integrals and related observables.

The derivation is completely general: it does not depend on whether the 
reference momenta are massless or massive, and it applies uniformly to all kinematic configurations considered in this work. For the special case of two-denominators, these relations were worked out explicitly in 
ref.~\cite{Lyubovitskij:2021ges}, and for the three-denominators, in refs.~\cite{Ahmed:2024pxr,Haug:2024yfi}.

\section{Conclusions}
\label{sec:concl}

In this work, we have computed four-denominator angular phase-space integrals within dimensional regularisation using the Mellin--Barnes (MB) representation. Our analysis covers all kinematic scenarios involving massless, single-massive, and multiple-massive momenta. For the massless and single-massive cases, we evaluated multi-fold MB integrals explicitly, obtaining results up to the finite order in the dimensional regulator in terms of Goncharov polylogarithms. The maximal transcendental weight of the GPLs appearing in these results is two. Both cases exhibit a single pole in the dimensional regulator.
For higher-mass configurations, we extended a partial fraction decomposition method, originally developed for two-propagator integrals, to the four-denominator case and provided explicit formulae. 

A key highlight of our work is the solution of six- and seven-fold MB integrals in terms of analytic functions such as GPLs---to our knowledge, a first in the literature. The power of this approach lies in its algorithmic and scalable nature, enabling potential applications far beyond the specific angular integrals considered here. While this method provides a systematic way to express results in terms of GPLs, it is not guaranteed that GPLs form the complete functional 
space for a given problem, as this depends on the perturbative order and mass scales involved. In such cases, the results can be expressed in terms of iterative integrals, which share several useful properties and constitute a highly practical representation. Similar situations occur for loop Feynman integrals.  

Finally, the angular part derived here can be systematically combined with the process-dependent radial or parametric component, making these results directly applicable to a broad class of multi-loop calculations and phase-space integrals in perturbative quantum field theory. While combining, one has to be careful in handling soft singularities, as explained in refs.~\cite{Ahmed:2024pxr,Ahmed:2024owh}. Since our results are expressed in terms of GPLs and evaluated in the physical scattering region where all $v_{ii} \in [0,1/4]$ and $v_{ij} \in [0,1/2]$, it is highly plausible that the parametric integrals can also be carried out analytically, yielding final results in terms of GPLs or other analytic functions. In cases where an analytic expression is not attainable, performing a one-dimensional parametric integral numerically still provides sufficient precision for phenomenological applications.

\section{Acknowledgments}
The work of TA and AR is supported by the Deutsche Forschungsgemeinschaft (DFG) through Research Unit FOR2926, \textit{Next Generation Perturbative QCD for Hadron Structure: Preparing for the Electron-Ion Collider}, project number 409651613. 

\bibliographystyle{JHEP}
\bibliography{main} 
\end{document}